\documentclass[a4paper,11pt]{article}
\pdfoutput=1
\usepackage{jcappub}
\usepackage{subfig}
\usepackage{epsfig}
\usepackage{graphicx}
\usepackage{color}
\usepackage{amssymb}
\usepackage{amsmath}
\usepackage{mathtools}
\usepackage{url}
\usepackage{natbib}
\usepackage[normalem]{ulem}
\usepackage{afterpage}
\usepackage{float}

\makeatletter
\let\jnl@style=\rm
\def\ref@jnl#1{{\jnl@style#1}}

\def\aj{\ref@jnl{AJ}}                   
\def\actaa{\ref@jnl{Acta Astron.}}      
\def\araa{\ref@jnl{ARA\&A}}             
\def\apj{\ref@jnl{ApJ}}                 
\def\apjl{\ref@jnl{ApJ}}                
\def\apjs{\ref@jnl{ApJS}}               
\def\ao{\ref@jnl{Appl.~Opt.}}           
\def\apss{\ref@jnl{Ap\&SS}}             
\def\aap{\ref@jnl{A\&A}}                
\def\aapr{\ref@jnl{A\&A~Rev.}}          
\def\aaps{\ref@jnl{A\&AS}}              
\def\azh{\ref@jnl{AZh}}                 
\def\baas{\ref@jnl{BAAS}}               
\def\bac{\ref@jnl{Bull. astr. Inst. Czechosl.}}
\def\caa{\ref@jnl{Chinese Astron. Astrophys.}}
\def\cjaa{\ref@jnl{Chinese J. Astron. Astrophys.}}
\def\icarus{\ref@jnl{Icarus}}           
\def\jcap{\ref@jnl{J. Cosmology Astropart. Phys.}}
\def\jrasc{\ref@jnl{JRASC}}             
\def\memras{\ref@jnl{MmRAS}}            
\def\mnras{\ref@jnl{MNRAS}}             
\def\na{\ref@jnl{New A}}                
\def\nar{\ref@jnl{New A Rev.}}          
\def\pra{\ref@jnl{Phys.~Rev.~A}}        
\def\prb{\ref@jnl{Phys.~Rev.~B}}        
\def\prc{\ref@jnl{Phys.~Rev.~C}}        
\def\prd{\ref@jnl{Phys.~Rev.~D}}        
\def\pre{\ref@jnl{Phys.~Rev.~E}}        
\def\prl{\ref@jnl{Phys.~Rev.~Lett.}}    
\def\pasa{\ref@jnl{PASA}}               
\def\pasp{\ref@jnl{PASP}}               
\def\pasj{\ref@jnl{PASJ}}               
\def\rmxaa{\ref@jnl{Rev. Mexicana Astron. Astrofis.}}%
\def\qjras{\ref@jnl{QJRAS}}             
\def\skytel{\ref@jnl{S\&T}}             
\def\solphys{\ref@jnl{Sol.~Phys.}}      
\def\sovast{\ref@jnl{Soviet~Ast.}}      
\def\ssr{\ref@jnl{Space~Sci.~Rev.}}     
\def\zap{\ref@jnl{ZAp}}                 
\def\nat{\ref@jnl{Nature}}              
\def\iaucirc{\ref@jnl{IAU~Circ.}}       
\def\aplett{\ref@jnl{Astrophys.~Lett.}} 
\def\apspr{\ref@jnl{Astrophys.~Space~Phys.~Res.}}
\def\bain{\ref@jnl{Bull.~Astron.~Inst.~Netherlands}} 
\def\fcp{\ref@jnl{Fund.~Cosmic~Phys.}}  
\def\gca{\ref@jnl{Geochim.~Cosmochim.~Acta}}   
\def\grl{\ref@jnl{Geophys.~Res.~Lett.}} 
\def\jcp{\ref@jnl{J.~Chem.~Phys.}}      
\def\jgr{\ref@jnl{J.~Geophys.~Res.}}    
\def\jqsrt{\ref@jnl{J.~Quant.~Spec.~Radiat.~Transf.}}
\def\memsai{\ref@jnl{Mem.~Soc.~Astron.~Italiana}}
\def\nphysa{\ref@jnl{Nucl.~Phys.~A}}   
\def\physrep{\ref@jnl{Phys.~Rep.}}   
\def\physscr{\ref@jnl{Phys.~Scr}}   
\def\planss{\ref@jnl{Planet.~Space~Sci.}}   
\def\procspie{\ref@jnl{Proc.~SPIE}}   

\usepackage{lmodern}

\usepackage[T1]{fontenc}
\usepackage[utf8]{inputenc}
\usepackage{color}
\usepackage[usenames,dvipsnames]{xcolor}
\usepackage{amsmath}
\usepackage{amssymb}
\usepackage{graphicx}
\usepackage{esint}

\newcommand{\half}{\frac{1}{2}}

\newcommand{\cinv}{\ensuremath{{\sf C}^{-1}}}

\newcommand{\bomu}{\ensuremath{\boldsymbol{\mu}}}

\newcommand{\boC}{{\ensuremath{\sf{C}}}}

\newcommand{\sfB}{\ensuremath{{\sf{B}}}}

\newcommand{\mathd}{\ensuremath{\mathrm{d}}}

\newcommand{\both}{\ensuremath{\boldsymbol{\theta}}}

\newcommand{\calP}{\ensuremath{\mathcal{P}}}

\newcommand{\calG}{\ensuremath{\mathcal{G}}}

\newcommand{\fatx}{\ensuremath{\boldsymbol{x}}}

\newcommand\spart{\;\raise1.0pt\hbox{$/$}\hskip-6pt\partial}

\title{Debiasing inference with approximate covariance matrices and other unidentified biases}

\author[a]{Elena Sellentin,}
\author[b]{Jean-Luc Starck}
\affiliation[a]{Leiden Observatory, Leiden University, Huygens Laboratory, Niels Bohrweg 2, NL-2333 CA Leiden, The Netherlands.}
\affiliation[b]{
AIM, CEA, CNRS, Universit\'e Paris-Saclay, Universit\'e Paris Diderot, Sorbonne Paris Cit\'e, F-91191 Gif-sur-Yvette, France.}

\emailAdd{sellentin@strw.leidenuniv.nl}

\abstract
{
When a posterior peaks in unexpected regions of parameter space, new physics has either been discovered, or a bias has not been identified yet. To tell these two cases apart is of paramount importance. We therefore present a method to indicate and mitigate unrecognized biases: Our method runs any pipeline with possibly unknown biases on both simulations and real data. It computes the coverage probability of posteriors, which measures whether posterior \emph{volume} is a faithful representation of \emph{probability} or not. If found to be necessary, the posterior is then corrected. This is a non-parametric debiasing procedure which complies with objective Bayesian inference.

We use the method to debias inference with approximate covariance matrices and redshift uncertainties. We demonstrate why approximate covariance matrices bias physical constraints, and how this bias can be mitigated. We show that for a Euclid-like survey, if a traditional likelihood exists, then 25 end-to-end simulations suffice to guarantee that the figure of merit deteriorates maximally by 22 percent, or by 10 percent for 225 simulations. Thus, even a pessimistic analysis of Euclid-like data will still constitute an 25-fold increase in precision on the dark energy parameters in comparison to the state of the art (2018) set by KiDS and DES. We provide a public code of our method.
}


\arxivnumber{}

\begin{document}
\maketitle
\flushbottom

\section{Introduction: unrecognized biases or new physics?}
The hardest mistakes to correct for, are those which remained unnoticed, or for which no solution exists yet. Contemporary cosmology actively tackles biases from covariance matrices \cite{SH15,SH17,TJK,Lacasa1}, likelihoods \cite{SHH18,Hahn}, lacking spectroscopic data for redshifts \cite{KiDS,Troxel}, and \cite{Kuijken2015} lists a comprehensive review of many more difficulties in leading weak lensing \cite{BartelSchneid} data analyses. 

Known and unknown biases propagate into cosmological parameter constraints, where they cause shifts of the posterior. In the absence of any biases, a posterior peaking in unexpected regions of parameter space must however be interpreted as a sign of new physics, and it is therefore of utmost importance to tell unrecognized biases and new physics apart. Furthermore, this distinction needs to be convincing beyond the boundaries of cosmology, i.e. also be convincing for neighbouring fields such as particle physics. 

We therefore here provide a method which safeguards cosmological parameter constraints against recognized or unrecognized biases. 

Based on a joint analysis of simulations and the real data with a likelihood, the method leads to unbiased credibility contours for the physical parameters. The method is non-Bayesian (but compatible with Bayesian inference) and therefore applies also when there is no error model available, which a Bayesian mitigation method would require. The thus gained credibility contours have a precise mathematical meaning, namely that of correct `coverage probability' (defined in section~\ref{meth}). Coverage probabilities of Bayesian posteriors objectively measure differences between frequentist and Bayesian parameter constraints.  They thereby measure how much the inferred physics depends on our \emph{assumptions} when analyzing the data, rather than on information contained \emph{in} the data. Accordingly, reporting the coverage also measures how much (frequentist) particle physicists, and (Bayesian) cosmologists could maximally disagree, given the same data set.

We develop our method in section \ref{meth}. The method is general, but was developed to address currently outstanding problems of cosmic shear. For example, \cite{SH15, SH17} derive the to-date only known completely bias-free likelihood for estimated covariance matrices. In \cite{SHInsuff}, it was then shown that extra-correlations exist between weak lensing data points, which cannot be captured by any covariance matrix, but affect the inference. In \cite{SHH18}, these extra-correlations were studied in detail, showing that the actual weak lensing likelihood must be skewed, and that this skewness translates into parameter biases up to 10 percent of the standard deviation, depending on how the weak lensing data are binned in angular ranges and redshift bins. 

A recurrent theme in these analyses was that weak lensing does not easily \cite{SLICS} lend itself to simulations, due to reacting to cosmic structures on the scale of galaxy groups, and due to these structures falling already into the strongly non-linear regime of structure formation. We therefore here seek to minimize the \emph{number} of simulations, thereby trading for high accuracy of the few simulations, and nonetheless gaining faithful parameter constraints from a joint analysis of data and simulations with a likelihood.

 Section \ref{debcov} mitigates parameter biases from approximate likelihoods, where our example uses approximate covariance matrices. Section \ref{debred} studies photometric redshift uncertaintites and shows that redshift uncertaintites alone (without biased redshifts) can be neglected in current weak lensing surveys. Section \ref{Euclid} shows that 25 end-to-end simulations of a Euclid-like \cite{DefStudyRep} survey, in conjunction with an independent likelihood for this survey, suffice to guarantee that the figure of merit deviates maximally by 22 percent from its optimum. For 225 end-to-end simulations, the figure of merit can be guaranteed to deteriorate by maximally 10 percent. As a result, it can be taken essentially for granted that the upcoming Euclid-like surveys will lead to an 25-fold increase in our knowledge of the dark energy equations of state parameters \cite{Fundamental} $w_0$ and $w_a$.

\section{Mitigating unrecognized biases: method and examples}
\label{meth}
To avoid that unrecognized biases feign new physics, we establish a method that takes as input any existing data analysis pipeline. The method runs the pipeline on simulations and real data alike, and then computes and corrects the \emph{coverage probability}. We describe why this procedure debiases parameter constraints.

\subsection{What do posteriors really measure?}
Biases in an inference cause that a posterior, or likelihood, exclude the true parameters too often, for example because the posterior is shifted or too narrow. The notion of `too often', is made mathematically precise by \emph{coverage probabilities}. The coverage probability of a posterior credibility contour is the fraction of times that this contour includes the true parameters, under repetitions of the experiment. The default expectation of most scientists is that the 68 percent credibility contour (as an example) contains the true parameters 68 percent of the time. In reality, however, the 68 percent posterior credibility contour is constructed such that it contains 68 percent of the \emph{posterior volume}. Most scientists expect that posterior volume measures (Kolmogorov) probability, but this is not necessarily so. We refer to this expectation by speaking of `correct coverage' for short \citep{CoverSyring,coverabc}. 

Coverage probabilities superficially sound like a frequentist concept, but so-called objective Bayesian analyses \cite{HSneutrinos} achieve the correct coverage probabilities as well \cite{Gruenwald,Rankcover}, due to their explicit construction of priors. Objective Bayesian analyses thus implement the correct noise propagation through mathematical derivations, with the result that posterior volume indeed measures probability. In contrast, the correct coverage is not automatically achieved by so-called subjective Bayesian analyses \cite{goldstein2006,Sub}. These regard priors as subject to choice, or use  hyper-parameters, approximate likelihoods \citep{SHH18,SHInsuff}, or idealized parametric models, with the result that the total Bayesian flow of information is not representative of nature, although mathematically self-consistent \citep{CoverSyring,Rankcover}.

In total, it cannot be taken for granted that posterior volume measures probability as expected, but such potential discrepancies can be \emph{reported} by quoting coverage probabilities. This is of direct relevance to tensions between experiments.

\subsection{Algorithm to measure the coverage probability of posteriors}
Any unrecognised or unintended systematic will affect the coverage. Hence, measuring the coverage can detect hidden biases, even if the source of the biases is unknown. Correcting the coverage is then a model-independent solution for mitigating unrecognized biases. We measure and correct the coverage probability with the following algorithm. 

A set of fiducial parameters $\both_0$ is chosen for $N$ simulations of artificial data sets $\fatx_i$, with $i \in [1,N]$. These simulations imitate the real data $\fatx_o$. We denote posterior \emph{densities} by curly capital $\calP$, and associated probabilities, that are scalar rather than densities, by roman $P$.

A state-of-the-art likelihood is then run on all simulations and also on the real data. This results in $N$ posteriors $\calP(\both|\fatx_i)$ from simulations, and the posterior $\calP(\both|\fatx_o)$ of the real data. For each of these $N+1$ posteriors, 120 credibility contours (or more) are computed. We provide a public code\footnote{Public at github.com/elenasellentin/Mitigate\_Unrecognized\_Biases}, where 100 of these contours are equidistant between zero and 99.9 percent posterior credibility. Twenty further contours are equidistant between 95.25 and 99.75 percent credibility. These finely spaced contours enable a reliable coverage correction in the outer tails of a posterior. If the data analysis pipeline contains biases, then the contours resulting from it will not cover correctly.

We denote by $\alpha$ fractions of the posterior volume, and accordingly $\alpha \in [0,1]$. We consider posterior contours that contain a fraction $\alpha$ of the posterior volume and which are isocontours of the posterior. They thus cut the posterior in a certain height below its peak. For each data set $i$, the posterior will be slightly differently shaped, and the height of the $\alpha$th contour thus changes with $i$. We therefore denote this height as roman $P^i_\alpha$, where $i \in [0,N]$ identifies the data set, and $\alpha \in [0,1]$ identifies the fraction of posterior volume that the contour contains.

Each of the posteriors will take a different (scalar) value \emph{at} the fiducial parameters of the simulation. We denote this value as roman $P(\both_0|\fatx_i)$, where the subscript zero indicates that this is the posterior probability assigned to the fiducial parameters.

The $\alpha$th credibility contour then contains the true parameters if the posterior value at the fiducial parameters is larger than the posterior height of the contour:
\begin{equation}
   P(\both_0|\fatx_i) \geq P^i_\alpha \Rightarrow \ \alpha\mathrm{th \ contour \ contains}\ \both_0.
\end{equation}We measure this for all contours, for all posteriors.
The coverage probability, $C_\alpha$, is then the probability $p$ that the $\alpha$-posterior credibility region contains (`covers') the true parameter values
\begin{equation}
C_\alpha = p (  \both_0 \ \mathrm{inside} \ \alpha\mathrm{th\ contour } ).
\label{eqcover}
\end{equation}
The default expectation  would be that $C_\alpha = \alpha$, meaning that posterior volume measures probability under repetition of the experiment. In contrast, if biases occurred in the analysis, then a credibility contour further out in the posterior will achieve coverage $C_\alpha$. 

For example, the allegedly 95 percent credibility contour of the biased analysis might be found to contain the true parameters only 90 percent of the times. Then it is in reality the 90 percent contour, until the bias is found and corrected. If the bias cannot be found, a mitigation is to discard the biased contours and instead adopt the contours of correct coverage. The new contours will then include the true parameters with the right fraction of times -- despite the bias being unknown.

\begin{figure}
\centering
\includegraphics[width=0.99\textwidth]{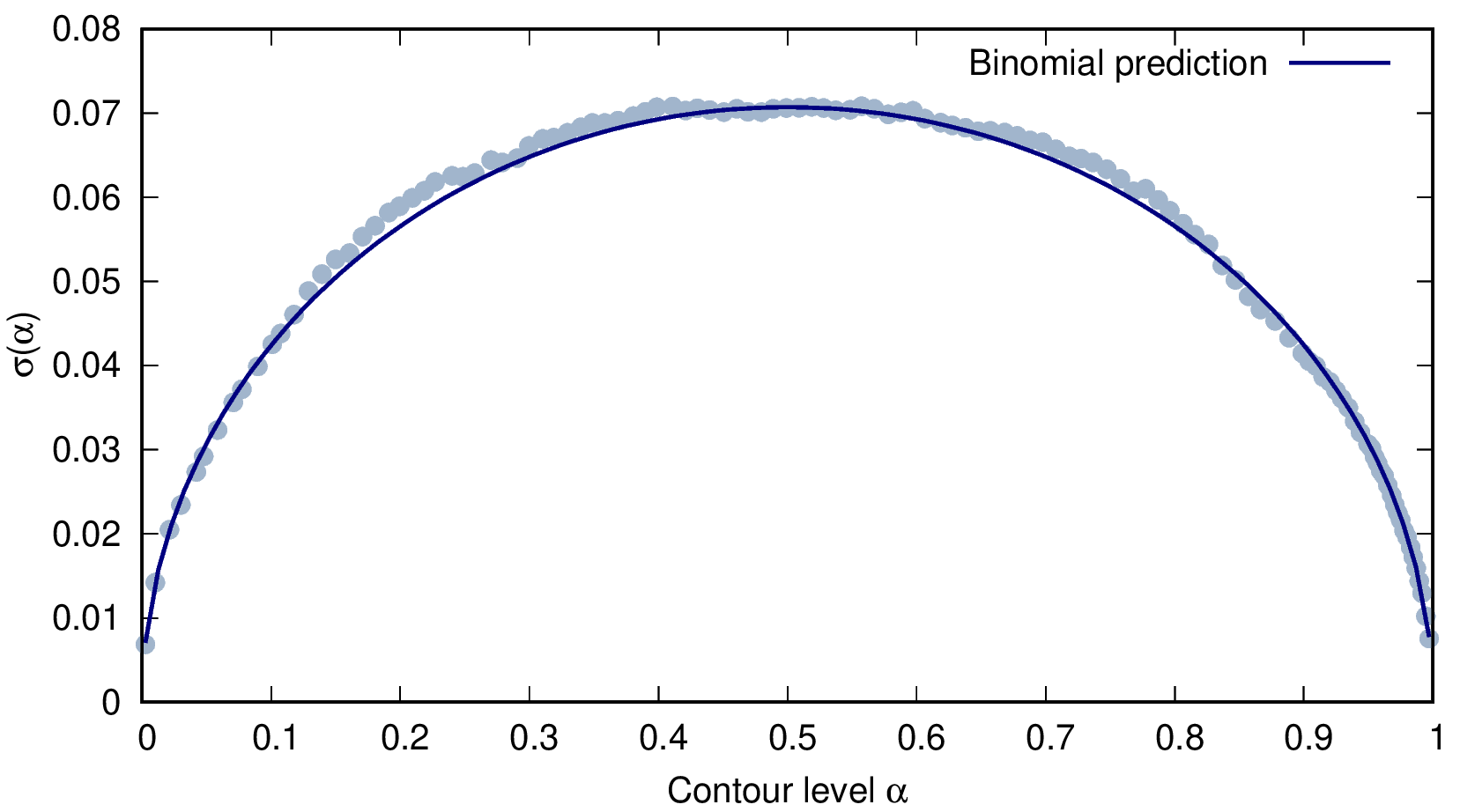} 
\caption{Standard deviation of the coverage estimator $\hat{C}_\alpha$, as a function of the contour level $\alpha$. The grey points indicate the numerically estimated standard deviations of a coverage measured from $N$ simulations, and the blue solid line indicates the prediction for the standard deviation from a binomial distribution.}
\label{fig:binomial}
\end{figure}

The coverage of equation \ref{eqcover} can be estimated from $N$ simulations, and we denote its estimator by $\hat{C}_\alpha$. This estimator simply counts how often the true parameters fall \emph{inside} the $\alpha$-contour. If they do not fall inside the contour, they fall automatically outside, and this either-or process indicates that the estimator $\hat{C}_\alpha$ must (by definition) follow a binomial distribution with success rate $\alpha$ and $N$ trials. The mean and standard deviation of the binomial distribution then give the mean and standard deviation $\sigma$ of our coverage estimator
\begin{equation}
    \langle \hat{C}_\alpha \rangle = C_\alpha, \ \ \ \mathrm{\sigma} = \frac{\sqrt{\alpha(1-\alpha)}}{\sqrt{N}}.
    \label{nerr}
\end{equation}
Figure~\ref{fig:binomial} shows that the binomial distribution models the noise in the estimated coverage correctly: for the innermost contours, where $\alpha$ is low, and for the outermost contours, the standard deviation is the smallest.

If credibility contours cover correctly, then $C_\alpha = \alpha$, and the standard deviations will in the following be adopted as error bars. 

We provide three simple examples of coverage correction in section \ref{examples}, before we apply the method to cosmological analyses in section \ref{cosmology}.

\subsection{Examples}
\label{examples}

\begin{figure}
\centering
\includegraphics[width=0.99\textwidth]{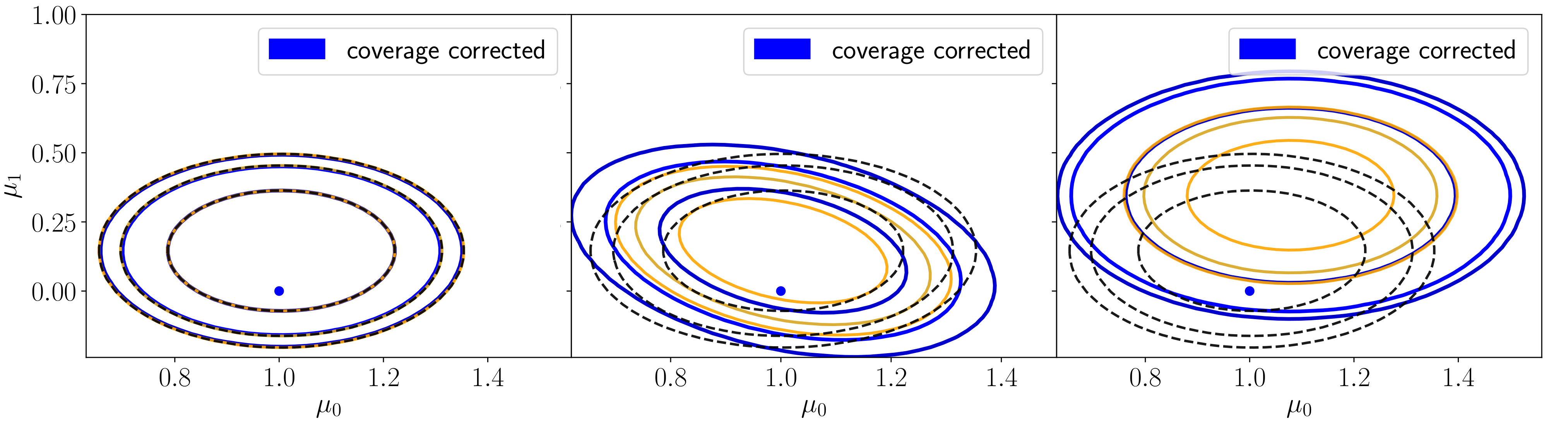} 
\caption{Examples of unrecognized biases corrected via coverage measurements. Dashed contours: the correct, bias-free analysis for comparison. Yellow: biased analysis per panel (left: no bias, middle: bias introduced by an approximate covariance matrix, right: bias introduced by an unintentionally informative prior). Blue dot: the true parameters, note how the yellow posterior in the right is so biased that it excludes this point. Blue contours: coverage corrected analysis, the true parameter point is now included with the right probability, even if the bias cannot be found. }
\label{fig:examples}
\end{figure}
Figure~\ref{fig:examples} illustrates three examples of mitigating unrecognized biases via coverage correction. The real data vector contains 100 data points, drawn from a Gaussian distribution with unit variance. The first fifty data points have mean $\mu_0 = 1$, the remaining data points have $\mu_1 = 0$. The parameters to be inferred are $p_0 = \hat{\mu}_0$ and $p_1 = \hat{\mu}_1$. We simulate 1000 artificial data vectors, by drawing from the same Gaussian. A bias is then introduced in the analysis, the coverage is measured and corrected, resulting in increased contour size.

Example 1, in the left panel of figure~\ref{fig:examples} is bias free: the data are analyzed with the correct Gaussian likelihood and a flat unbounded prior, which produces automatically the correct coverage for linear parameters, such as $\mu_1, \mu_2$. Example 2, in the middle panel, analyzes the data with a biased inverse covariance matrix. The correct inverse covariance would have been $\boC^{-1} = \mathbb{I}$, the identity matrix, but the off-diagonal elements were changed to $C^{-1}_{ij} = 1e^{-2}$. Section \ref{cosmology} will detail why biased covariance matrices shift posteriors, here we only illustrate that our method is able to correct for this, without needing to know the origin of the error.

\begin{figure}
\centering
\includegraphics[width=0.89\textwidth]{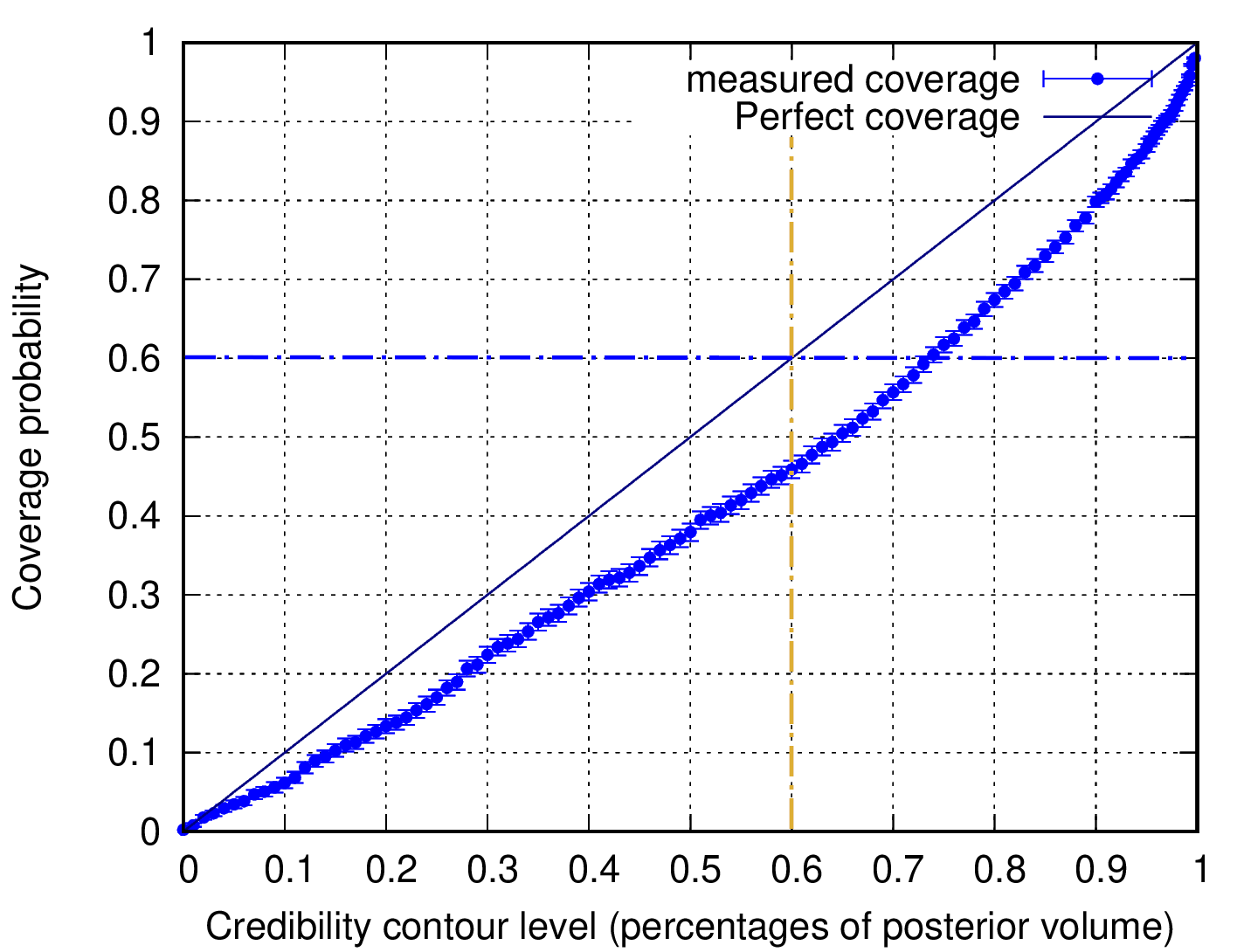}
\caption{Measured coverage probabilities of example 2, the middle panel of figure~\ref{fig:examples}, where the bias is due to an approximate covariance matrix. Horizontal axis: the credibility contours, computed as percent of (biased) posterior volume, where the posterior is too narrow. The vertical axis shows the coverage probabilities of the contours, this is the probability that a given contour includes the true parameter values. The diagonal line is a bias-free analysis, where posterior volume reflects probability. However, due to the bias of example 2, all contours contain the true parameters systematically fewer times than expected (they are too narrow). For example, the 60\% credibility contour contains 60\% of posterior \emph{volume} (yellow cut), but it contains the true parameter point only 48\% of the times. After measuring the coverage, this can be corrected: the blue cut shows that the (biased) 72\% posterior volume contour contains the true parameters 60\% of the time. The (biased) 72\% volume contour is thus the (unbiased) 60\% confidence contour. The unknown bias is thus mitigated by relabelling the contours.}
\label{fig:cover2}
\end{figure}

Figure~\ref{fig:cover2} plots the measured coverage probabilities of example 2. Due to the hidden bias, the contours are systematically to small, resulting in the seen undercoverage of figure~\ref{fig:cover2}.

Finally, example 3 corrects the effects of an unintentionally informative prior $\pi$, given by
\begin{equation}
    \pi(p_0,p_1) = \calG(p_0|\mu=1.45,\sigma^2=0.1)\calG(p_1|\mu=1.35,\sigma^2=0.1),
\end{equation}
where $\calG$ is the Gaussian distribution. The prior is so informative that the biased posterior excludes the true parameters (blue point). After coverage correction, the true parameters are again included. Plotted contours lie at 68, 90 and 95 percent posterior volume (before coverage correction, yellow), and at 68, 90 and 95 percent coverage probability (after correction, blue).

\subsection{Blind spots of the method}
The method detects discrepancies between simulations and the assumptions of a data analysis pipeline. It then corrects for these discrepancies when analyzing the real data. Consequently, it cannot correct for systematic effects which are omitted in both simulations and the analysis pipeline. For example, if neither a likelihood, nor the simulations include a survey mask, then the method cannot correct for imperfections in survey mask handling. Likewise, if the simulations implement precisely the same assumptions as the analysis pipeline, then a self-confirming situation is created, which the method also cannot detect. If the posterior then peaks nonetheless in unexpected regions, then the real data obey other physical or statistical laws than the ones simulated.

If the simulations lack in accuracy, then our method suffers from the same difficulties as any simulation-based inference technique. We shall however find in section~\ref{Euclid} that our method requires by many orders of magnitude the fewest number of simulations \cite{TJK,SH17}. This arises due to the the joint analysis with a likelihood, which already contains statistical information which simulations would otherwise need to provide.

\section{Applications to cosmology}
\label{cosmology}
In this section we apply coverage calibration to cosmic shear (weak lensing) analyses \cite{KiDS,Troxel, BartelSchneid}, where approximate covariance matrices and redshift uncertaintites often introduce biases of unknown magnitude and of unknown parametric form.

\subsection{Why approximate covariance matrices shift posteriors}
Approximate covariance matrices are today used in weak lensing \citep{KiDS,Troxel}, but also supernova analyses adapt their covariance matrices to achieve a desired goodness of fit \citep{PerlmutterSN,RiessSN}. One often encountered preconception is that such approximate covariance matrices only affect the width of posterior contours, but not where a posterior peaks. We therefore explain why the opposite is true: We show that using an approximate covariance matrix is mathematically the same as fitting to a biased data set, and systematic parameter shifts will ensue.

Consider a Gaussian likelihood, as is currently standard in cosmology
\begin{equation}
\calG(\fatx, \bomu, \boC  ) = \frac{1}{\sqrt{ (2\pi)^p |\boC| }} \exp \left( -\half (\fatx-\bomu)^\top \cinv (\fatx - \bomu)\right),
\label{gauss}
\end{equation}
where, $\fatx$ is a $p$-dimensional data vector and the superscript $\top$ denotes transposition. The mean $\bomu(\both)$ is a function of the parameters $\both$ to be inferred, and the covariance matrix is $\boC$. Parameters are then estimated by sampling the posterior, which is the likelihood times a prior.

To isolate the effect of approximate covariance matrices, we assume unbounded flat priors, and that the data $\fatx$ contain no systematic effects. Such sound data are then nonetheless \emph{effectively} transformed into a biased data set, if an approximate covariance matrix is used in the analysis. This can be seen as follows.

Let the correct covariance matrix be $\boC_{\rm c}$ and let $\boC_{\rm B}$ be an approximation of it. Both are symmetric positive-definite matrices.

If an analysis uses the correct covariance matrix, the best fit lies where
\begin{equation}
\chi^2_{\rm c} = (\fatx - \bomu_{\rm c})^\top \cinv_{\rm c} (\fatx - \bomu_{\rm c}),
\label{truth}
\end{equation}
is minimal. Equation~\ref{truth} describes that the parameters of the model $\bomu$ will adjust to minimize the distance to the data $\fatx$. The best-fitting parameters are then $\both_{\rm c}$ for which $\bomu_{\rm c} = \bomu(\both_{\rm c})$. During minimization, statistical compatibility between the mean and the data is measured in units of the inverse covariance matrix. If we exchange the covariance matrix, this distance measure changes. In the units of the new covariance matrix, another $\bomu(\both)$ will then be closest to the data $\fatx$. Consequently, the parameters $\both$ will adapt, in order to produce this new mean as well as possible.

We now relate the two matrices via the function
\begin{equation}
\cinv_{\rm B} = \sfB^\top \cinv_{\rm c} \sfB,
\label{rot}
\end{equation}
where a bias occurs if $\sfB \neq \mathbb{I}$. The left- and right-multiplication by $\sfB$ is convenient, but not a specialization. We could equally have written
\begin{equation}
\cinv_{\rm B} = \cinv_{\rm c} + {\sf \Delta},
\label{add}
\end{equation}
where ${\sf \Delta}$ is the matrix of additive inaccuracies. Since $\cinv_{\rm c}$ is a symmetric matrix, ${\sf \Delta}$ is also by construction symmetric. The matrix ${\sf B}$ is then guaranteed to exist, since for symmetric matrices ${\sf A}$ any congruent matrix $\sfB^\top {\sf A} \sfB$ is again symmetric for all $\sfB$, and equations~\ref{rot} and~\ref{add} are both valid ways of describing the systematic uncertainties in a covariance matrix. The corresponding additive uncertainty is then
\begin{equation}
{\sf \Delta} = \sfB^\top \cinv_{\rm c} \sfB - \cinv_{\rm c}.
\end{equation}

Since $\cinv_{\rm c}$ is unknown, cosmology is forced to use $\cinv_{\rm B}$ for the likelihood. The thus gained $\chi^2$-squared surface is then minimized where
\begin{equation}
\chi^2_{\rm B} = (\fatx - \bomu_{\rm B} )^\top (\sfB^\top \cinv_{\rm c} \sfB) (\fatx - \bomu_{\rm B}),
\label{biased}
\end{equation}
is minimized. This occurs at a new mean $\bomu_{\rm B} = \bomu(\both_{\rm B})$, and the parameters $\both_{\rm B}$ will differ from $\both_{\rm c}$. 

If we conduct a thought experiment where we forget that the new parameters differ, we see that using a biased covariance matrix is akin to analyzing a biased data set $\fatx_{\rm B}$ with the correct covariance matrix. To see this, we set $\bomu_{\rm B} = \bomu_{\rm c}$ in our thought experiment. Then, to yield as good a best fit as when using the correct covariance matrix, we have to demand
\begin{equation}
\sfB(\fatx_{\rm B} -\bomu_{\rm c}) = (\fatx - \bomu_{\rm c}).
\end{equation}
This can be solved for $\fatx_{\rm B}$, and we find
\begin{equation}
\fatx_{\rm B} = \sfB^{-1} \left[ \fatx + \bomu_{\rm c}(\sfB - \mathbb{I})\right].
\label{xb}
\end{equation}
This shows that using an incorrect covariance matrix $\sfB^\top \cinv_{\rm c} \sfB$ to analyze a sound data set is mathematically equivalent to analyzing the biased data set $\fatx_B$ with the correct covariance matrix. Only if $\sfB$ equals the identity matrix does $\fatx_{\rm B}$ coincide with $\fatx$.

In cosmology, the data $\fatx$ are of course fixed. The only free variables to compensate for the bias in the covariance matrix are then the cosmological parameters $\both$. The incorrect covariance matrix will consequently force the likelihood to peak at biased parameter values. 

In fact, in order for the biased equation~\ref{biased} to reproduce as good a fit as the correct equation~\ref{truth}, the relation
\begin{equation}
\sfB(\fatx-\bomu_{\rm B}) = \fatx - \bomu_{\rm c},
\label{linmap}
\end{equation}
needs to hold. Solving for the now preferred $\bomu_{\rm B}$, we find
\begin{equation}
\bomu_{\rm B} = \fatx - \sfB^{-1} ( \fatx - \bomu_{\rm c}).
\label{central}
\end{equation}
The parameters $\both$ will then attempt to create the mean $\bomu_{\rm B}$ instead of the mean $\bomu_{\rm c}$. Depending on the flexibility of the model, the parameters may not fully succeed in this. In total, we see however that a shift in parameters will ensue, and the direction and magnitude of the shift depends on the drawn data vector $\fatx$, and the biasing matrix $\sfB$, according to equation~\ref{central}.

\subsection{Undetectability in Fisher matrix forecasts}
The effect of approximate covariance matrices biasing parameters is invisible in Fisher matrix forecasts \cite{TegHeavTay,DefStudyRep}, because on average, one has $\langle \fatx\rangle = \bomu_{\rm c}$, and equation~\ref{central} then predicts $\bomu_{\rm B} = \bomu_{\rm c}$.  Fisher forecasts will therefore underestimate the total uncertainty. The biasing effect will only occur when analysing real data, where $\fatx$ is fixed to the realization on the sky. Equation~\ref{central} then describes that noise can be incorrectly interpreted as `signal' when the wrong covariance matrix is employed. In the following section we will present an example of thus resulting parameter biases.


\subsection{Forcing the KiDS-450 data to prefer the Planck cosmology}
\label{force}
Concerning how approximate covariance matrices bias physical parameters, we here illustrate that direction and magnitude of the posterior shift can also be controlled. Additionally, the goodness of fit can also be kept constant. A reduced-$\chi^2$ of order unity is therefore by no means a good indication that the best-fitting parameters are unbiased.

We illustrate this for the public KiDS-450 data from \cite{KiDS}, and force these data to prefer the Planck cosmology. DES \cite{Troxel} analyses could equally have been used. We underline that we here \emph{force} this transition to the Planck best-fitting cosmology. The aim of this study is dual, namely first to understand which data points are affected, and secondly to understand which procedures must be put in place in order to prevent such shifts.

We work with the original KiDS-450 data vector of 130 elements, which are the real-space estimators $\xi_+$ and $\xi_-$ \cite{BartelSchneid,Kilb} in four tomographic redshift bins and their cross-bins. Our weak lensing setup to compute the theory vector $\bomu(\both)$ is identical to \cite{KiDS} and our code has been verified against the code of \cite{KiDS}, leading to identical results for the theory vectors, given identical input parameters. We use \textsc{CLASS} \citep{Class1, Class2}, and \textsc{Halofit} \citep{HaloCamb3} for the non-linear power spectrum. We fix the spectral index $n_s$ and the reduced Hubble constant $h$, to Planck-motivated values of $h = 0.678$ and $n_s = 0.96$. Varying the cold dark matter density $\Omega_m$ and the normalization of the power spectrum $\sigma_8$, we find the best-fitting cosmology for the KiDS-450 data vector when analyzed with the public KiDS-450 covariance matrix to be
\begin{equation}
\Omega_m = 0.2,\ \sigma_8 = 0.838,
\end{equation}
with a $\chi^2 = 202$\footnote{The high value of this $\chi^2$ results from having fixed $n_s$ and $h$ to the Planck best-fitting values, rather than the KiDS best-fitting values.}. By transforming the covariance matrix, we now force the KiDS-450 data to prefer the Planck cosmology. This can be repeated for arbitrarily many parameters. 

We precompute the cosmological predictions $\bomu(\both)$ on a grid, and store the results, in order to make the upcoming analyses of this paper numerically feasible.

\begin{figure*}
\includegraphics[width=0.98\textwidth]{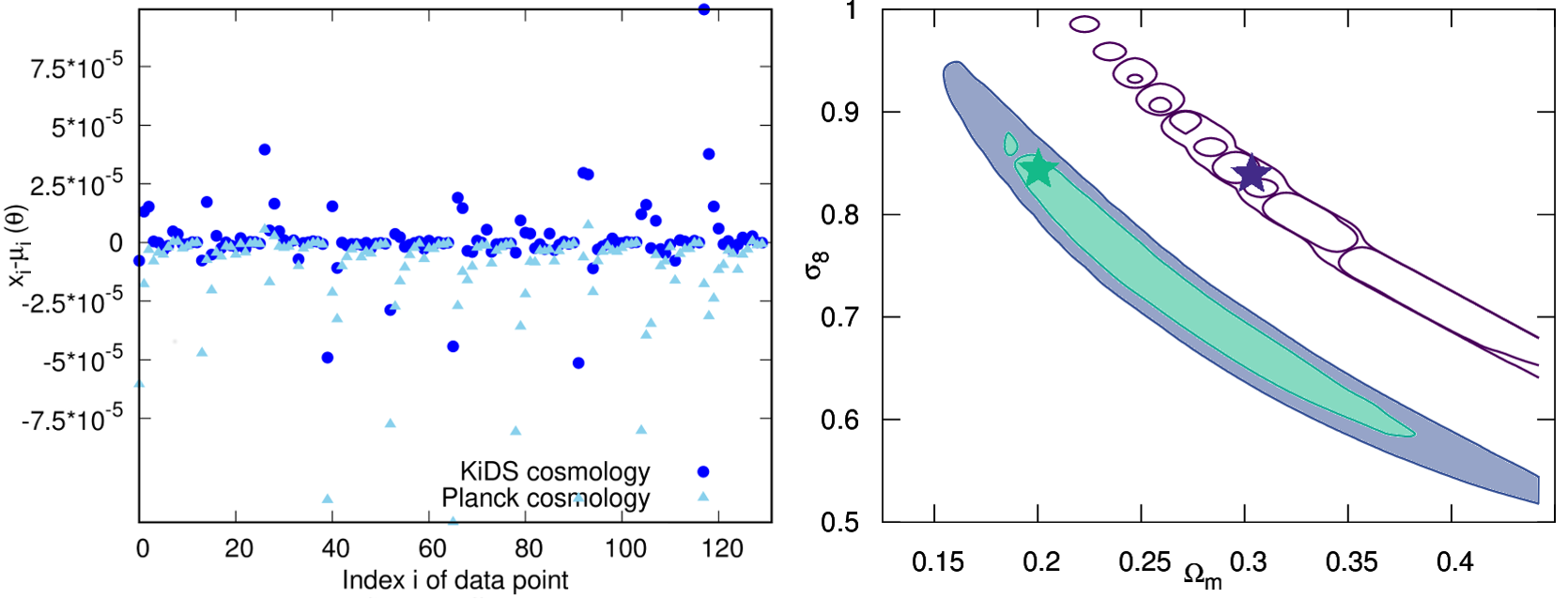} 
\caption{Left: The differences $\fatx-\bomu(\both)$ for the KiDS best-fitting cosmology (blue dots), and the Planck best-fitting cosmology (triangles). To achieve a translation of the posterior from the original KiDS best-fit (located at the green star in the right plot) to the Planck best-fit (located at the purple star in the right plot), a covariance matrix must be constructed which expects the noise pattern on the left as an indication of `strong covariance'. Right: The posterior with solid contours uses the original KiDS data and the original KiDS covariance matrix. The posterior in open contours uses the original KiDS data and the newly constructed covariance matrix. Its deformed shape results from having changed the determinant of the covariance matrix. The contours lie at 68\% and 90\% of posterior volume (which is the standard procedure in cosmology). }
\label{Posterior}
\end{figure*}

\begin{table}
\centering
\caption{Data points of KiDS-450 which are most unstable with respect to noise reassessment. The data point identifier `\#' counts from 1 and has the KiDS-ordering. The angular cuts in the corresponding DES data \cite{Troxel} exclude these data points. Together with \cite{SHInsuff}, the effect now repeatedly occured that data vector truncation influences the physical parameter constraints, which motivates that blinding strategies should be kept for future analyses. }
\label{Tab0}
\begin{tabular}{|lcr||lcr|} 
\hline
\# & angle (arcmin) &  $\xi_\pm$ & \# & angle (arcmin) & $\xi_\pm$ \\
\hline 
66	& 0.713  & $\xi_+$ & 92	& 0.713  & $\xi_+$ \\
40	& 0.713  & $\xi_+$ & 53	& 0.713 & $\xi_+$ \\
105	& 0.713 & $\xi_+$ & 79	& 0.713 & $\xi_+$ \\
1	& 0.713 & $\xi_+$ & 14	& 0.713 & $\xi_+$  \\
42	& 2.956 & $\xi_+$  & 54	& 1.452 & $\xi_+$  \\
67	& 1.452 & $\xi_+$ & 80	& 1.452 & $\xi_+$\\
106	& 1.452 & $\xi_+$ & 107	& 2.956 & $\xi_+$ \\
119	& 1.452 & $\xi_+$ & 120	& 2.956 & $\xi_+$ \\
95	& 6.017 & $\xi_+$ & 16	& 2.956 & $\xi_+$ \\
28	& 1.452 & $\xi_+$ & 41	& 1.452 & $\xi_+$ \\
		\hline
	\end{tabular}
\end{table}

We use a Planck best-fitting cosmology with \citep{Planck2015}
\begin{equation}
\begin{aligned}
n_s = 0.96,&\ h = 0.678\\
\Omega_m = 0.308,&\ \sigma_8 = 0.83.\\
\end{aligned}
\end{equation}
The original KiDS-450 analysis \citep{KiDS} leads to posterior constraints on $\sigma_8$ and $\Omega_m$ which are in tension with the Planck constraints. The left panel of figure~\ref{Posterior} plots the result of subtracting the KiDS-best fitting cosmology, or the Planck best-fitting cosmology from the KiDS data vector. Subtracting the Planck best-fitting cosmology leads to multiple sequences of adjacent data points being systematically below the mean (blue triangles in the negative domain). A covariance matrix can be tricked into expecting such a situation: By definition we have that the covariance between data points $x_i$ and $x_j$ is
\begin{equation}
c_{ij} = \mathbb{E}(x_ix_j) - \mathbb{E}(x_i) \mathbb{E}(x_j),
\label{expect}
\end{equation}
where $\mathbb{E}$ denotes taking the expectation value. Since this is an expectation value, a covariance matrix does not simply describe noise, but is rather extremely prescriptive: a positive covariance between data point $x_i$ and $x_j$ describes that \emph{if} data point $x_i$ is below the mean, then data point $x_j$ is \emph{expected} to be below the mean as well. We can hence construct a covariance matrix that expects the noise pattern of the blue triangles in figure~\ref{Posterior} and interprets it as a strong positive correlation between all data points that are below the mean. The data points whose noise will thereby be most strongly reassessed are listed in table \ref{Tab0}, which illustrates that it is consistently the estimators $\xi_+$ on the lowest angular scales (mostly $0.71$ and $1.45$ arcmins) who will show instability with respect to cosmological parameters, when their noise is reassessed. In this context it is interesting to note that the DES survey \cite{Troxel} excludes $\xi_+$ on such low scales, which will be partially responsible for why DES posteriors are closer to Planck than KiDS-450 posteriors.

We denote the original KiDS-450 covariance matrix as $\boC_{\rm KiDS}$. The minimal $\chi^2$ is then reached for
\begin{equation}
\chi^2 = 202 \ {\rm for\ } \boC_{\rm KiDS}, {\rm \ at \ } \both_{\rm c} = (0.2,0.838, 0.678,0.96),
\end{equation}
where the parameter vector is ordered as $\both = (\Omega_m, \sigma_8, h, n_s)$. We now demand that the KiDS data vector instead produces the Planck cosmology $\both_{\rm B} = (0.308, 0.83, 0.678, 0.96)$ as best fit, and solve for the matrix $\sfB$ from equation~\ref{central} which enables this.

\begin{figure*}
\includegraphics[width=0.99\textwidth]{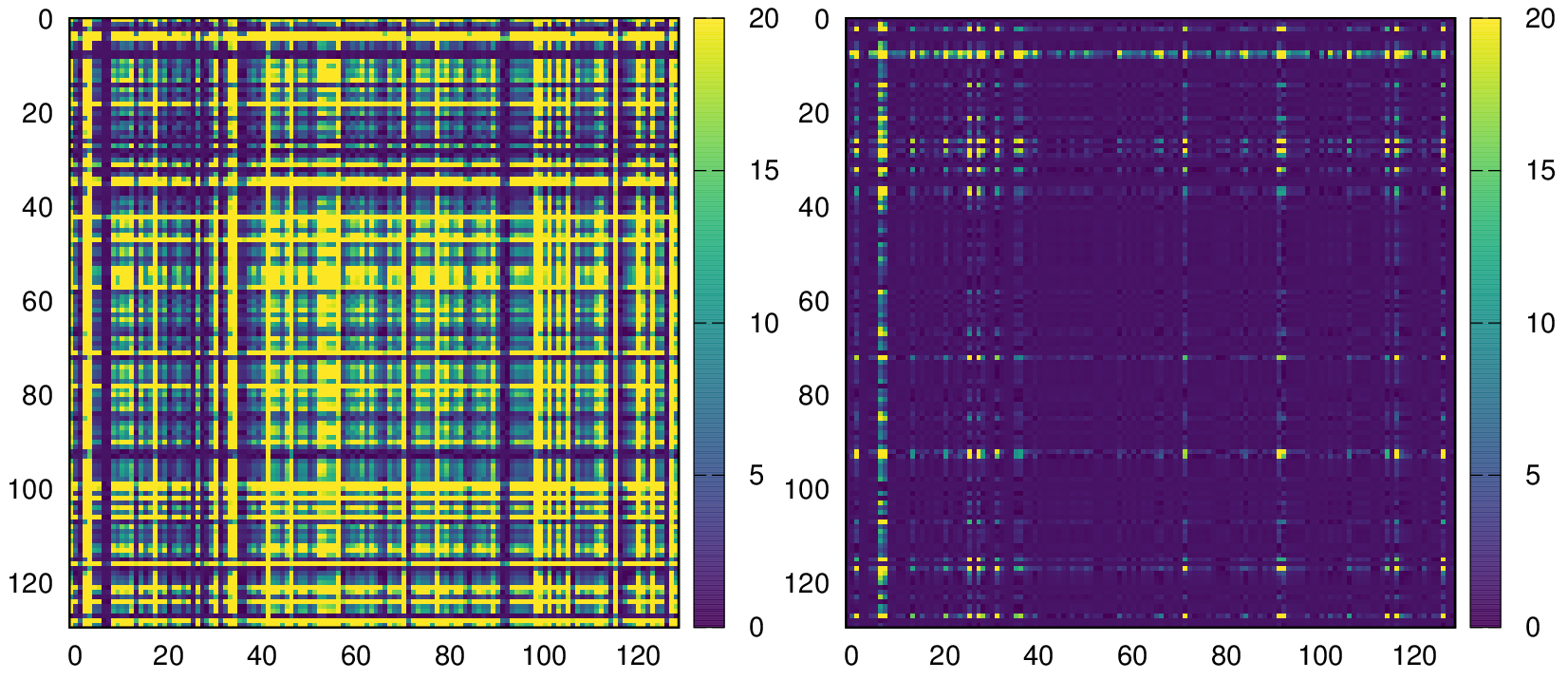} 
\caption{Left: Relative difference between the KiDS covariance matrix $\boC$, and the transformed matrix $\sfB^{-1} \boC \sfB^{-1}$. Right: As left, but now for the inverse covariance matrix. Changes in only a few columns of the precision matrix are sufficient to induce major biases in the physical parameters.}
\label{KiDS2Planck}
\end{figure*}

Since the matrix $\sfB$ has $p \times p$ entries, but equation~\ref{central} only poses $p$ constraints, reconstructing $\sfB$ is an under-determined system. There will hence be infinitely many solutions for $\sfB$, which directly implies that trying to debias an approximate covariance matrix is bound to fail.

Here, we now  pick out one solution, by demanding $\sfB$ to be diagonal, $\sfB = {\rm diag}(b_1,b_2,...,b_p)$. The required diagonal elements to force the KiDS data to prefer the Planck cosmology then follow to be
\begin{equation}
b_n = \frac{(\fatx - \bomu_{\rm c})_n}{(\fatx -\bomu_{\rm B})_n} \ \ \forall n \in [1,p].
\label{study}
\end{equation}
Using the original KiDS-450 data vector, and transforming the inverse KiDS covariance matrix to $\cinv_{\rm KiDS} \rightarrow \sfB^\top\cinv_{\rm KiDS} \sfB$, the Planck cosmology indeed becomes the new best fit
\begin{equation}
{\rm\ KiDS\  data,\ \sfB^\top\boC^{-1}_{\rm KiDS} \sfB:\ } \Delta \chi^2 = 0 {\rm \ at \ } \both_{\rm Planck}.
\end{equation}
The two posteriors arising from analyzing the KiDS data with the two covariance matrices are depicted in figure~\ref{Posterior}. This figure illustrates the successful translation of the posterior, although data and physical model were not changed. The Planck cosmology now fits the KiDS data with the same goddness of fit (the same $\chi^2$) as the KiDS best-fitting cosmology fitted the KiDS data before. Also visible is, however, that the new posterior is deformed. This side effect arises because the determinant of the covariance matrix was changed\footnote{Keeping the determinant constant would impose only one additional constraint, still leading to infinitely many solutions for $\sfB$, again leading to the conclusion that a non-parametric method is needed to debias inference with approximate covariance matrices.}.

We compute the relative differences between the original and the transformed covariance matrices. The matrix of relative differences is given by
\begin{equation}
R_{ij} = \frac{|C_{ij} - \tilde{C}_{ij}|}{|C_{ij}|},
\end{equation}
where $i$ and $j$ are the indices of the matrix elements and $\tilde{\boC}$ is shorthand for the transformed matrices. The left panel of figure~\ref{KiDS2Planck} shows the relative difference matrix $R_{ij}$ for the transformed  covariance matrix $(\sfB^{-1})^\top \boC_{\rm KiDS}\sfB^{-1}$, and the right panel shows the relative difference matrix $R_{ij}$ of the transformed inverse covariance matrix $\sfB^\top \cinv_{\rm KiDS}\sfB$.  The difference between the left and the right panel highlights the unpredictability of the matrix inversion: even if most columns in ${\sf C}_{\rm KiDS}$ are drastically changed, these changes can be redistributed during the inversion, and it is thus important to judge the accuracy of an \emph{inverse} covariance matrix directly.

 Figure~\ref{KiDS2Planck} reveals factor 20 changes in certain elements of the inverse covariance matrix. This is to be compared to the DES reanalysis \cite{DESdoesKiDS} of KiDS-450, where the reanalysis implemented factor 3 changes in the shape noise contribution to elements of the covariance matrix, and parameter shifts were found. We therefore conclude that a debiasing procedure for approximate covariance matrices is indeed needed.

As infinitely many solutions exist to induce a bias such that any arbitrary cosmology becomes the best-fitting cosmology, a parametric Bayesian treatment will not be able to debias the inference. In the following section we will hence reverse the workflow, and accept that any fixed covariance matrix of unknown bias will necessarily be used, and we debias the thus resulting parameter inference with coverage calibration.

\subsection{Debiasing inference with approximate covariance matrices of unknown bias}
\label{debcov}
In this section, we illustrate how to compute unbiased credibility contours for cosmological parameters, despite using a covariance matrix of unknown but non-zero bias.

 A necessary prerequisite for our method are $N$ accurate simulations. Importantly, these $N$ simulations are not used to compute a covariance matrix, or its inverse -- they are used to debias the inference pipeline which uses the approximate, analytical covariance matrix. To compute a numerical covariance matrix from simulations, one would require $N \gg p$, where $p$ is the dimension of the data set. To run our debiasing procedure, significantly fewer simulations are needed, and their number does not scale with the dimension of the data set either, see equation~\ref{nerr} and section~\ref{Euclid}.

We again use KiDS-450 as an example. For current weak lensing studies, sufficiently many or accurate simulations do not yet exist to conduct a coverage measurement. KiDS-450 posesses 930 simulations for 100 square degree sky patches \cite{Harno}, but spans by itself approximately 450 square degree. DES uses 18 simulations in \cite{MacCrannSims}, where the number of simulations is now the limiting factor. 

To demonstrate our method, we therefore generate 100.000\footnote{This large number resulted from experimenting with run time constraints. Far fewer are needed in reality, see section \ref{Euclid} for Euclid requirements.} Gaussian realizations of data vectors with the KiDS best-fitting cosmology as mean, and with the public KiDS covariance matrix. These shall serve as our simulations replacement.

\begin{figure*}
\includegraphics[width=0.98\textwidth]{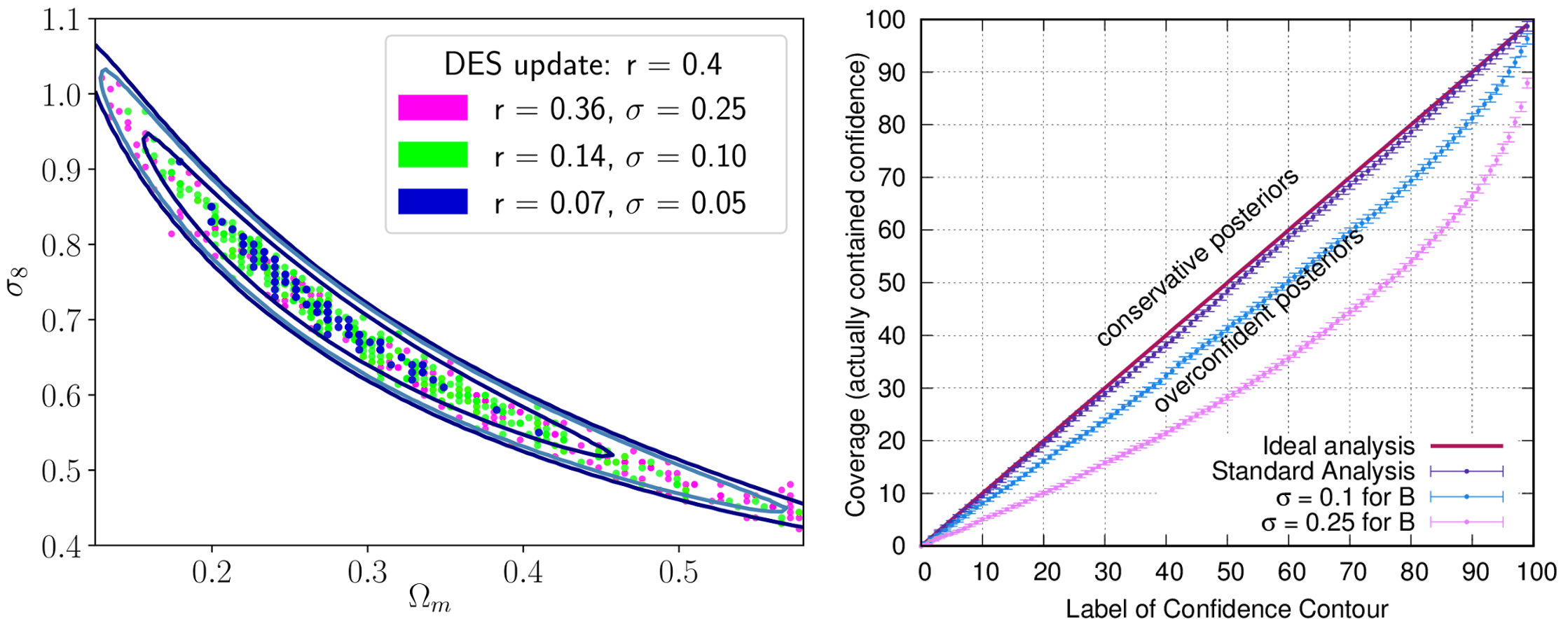} 
\caption{Left: The original KiDS posterior is depicted in open contours. The dots indicate how the best-fit scatters around if the KiDS data set is analyzed with differently biased covariance matrices. The entire posteriors shift along with the new best-fit, but the shifted contours are not shown for reasons of plot overcrowding. The relative biases $(r=0.07,0.14,0.36)$ here introduced to the covariance matrix are smaller than the changes applied by the DES reanalysis of KiDS ($r=0.4$) \cite{DESdoesKiDS}. Right: Measured coverages of the biased posteriors. The more biased the inverse covariance matrix is, the more does the biased posterior undercover (meaning it is too narrow). The posterior is debiased in figure \ref{Coverplot}.}
\label{CoverCalKiDS}
\end{figure*}

\begin{figure*}
\includegraphics[width=0.9\textwidth]{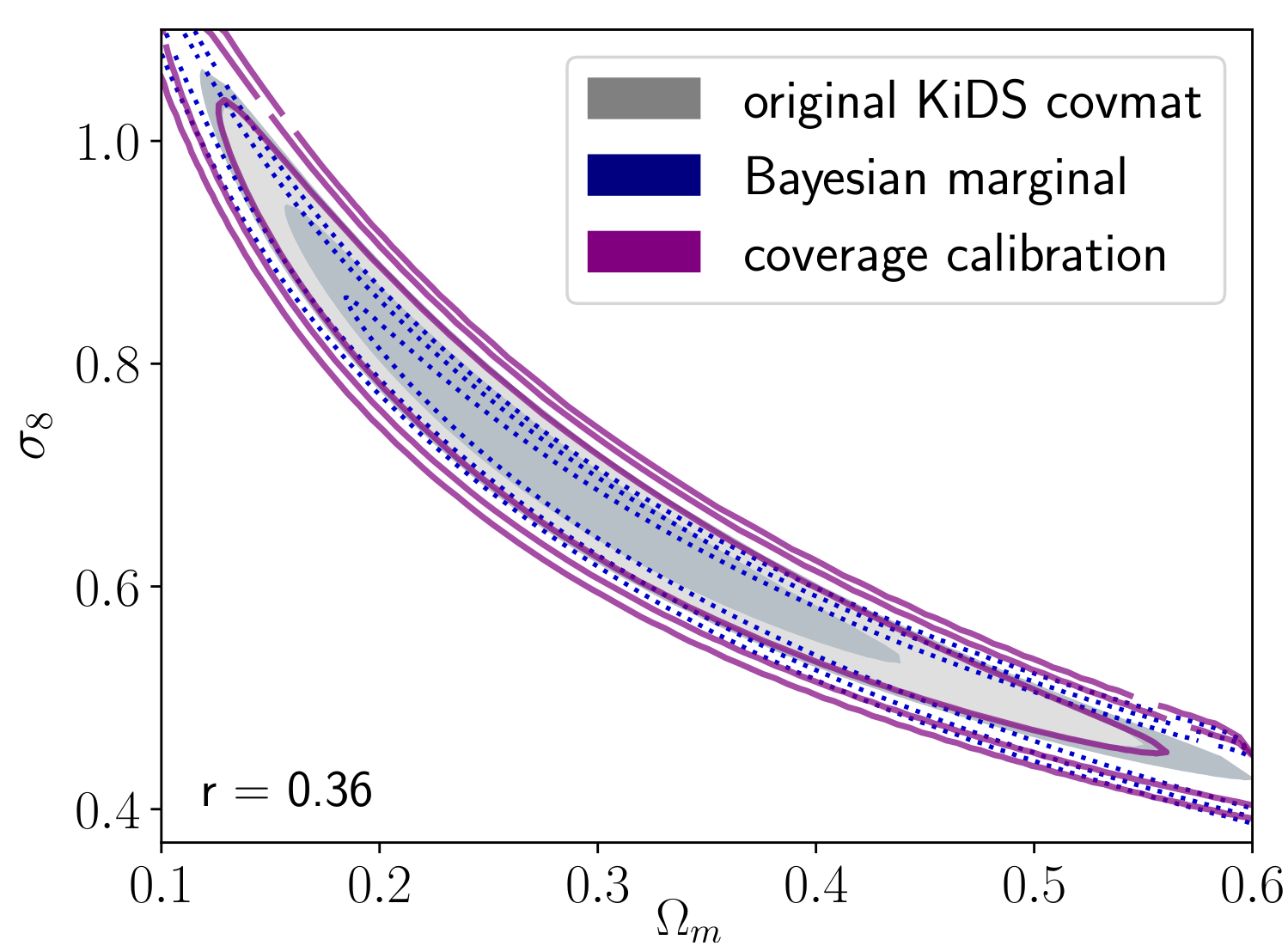}
\caption{KiDS posteriors, with and without propagation of covariance matrix uncertainty. This plot uses $\sigma = 0.25$ which leads to relative changes of $r = 0.36$ in the covariance matrix. Given \cite{DESdoesKiDS}, these are realistic values for current covariance matrix uncertainties. Solid contours: original KiDS-450 posterior, without propagating covariance matrix uncertainty. Blue dotted: Propagating the uncertainty via Bayesian marginalization, here possible since the toy-model for the bias is known, but impossible in reality where the model is unknown. Purple open contours: debiasing the credibility contours via coverage calibration. Comparison of solid grey and purple contours: for $r = 0.36$, the $68\%$ credibility contour of the debiased posterior is as large as the $90\%$ posterior credibility contour of the biased posterior (grey).}
\label{Coverplot}
\end{figure*}

We then precompute the theory vectors $\bomu(\both)$ on a grid in the $\Omega_m, \sigma_8$-plane, and then compute the 100.000 posteriors. The posterior per data vector $\fatx$ is
\begin{equation}
\calP(\both|\fatx) \propto L(\fatx|\both) \pi(\both),
\end{equation}
where $L(\fatx|\both)$ is the Gaussian likelihood, and $\pi$ are priors on the parameters.
We use top-hat priors, with 
\begin{equation}
0.09 < \Omega_m < 0.65,\ \ \ 0.37 < \sigma_8 < 1.1.
\end{equation}
Finally, the coverage is computed.

 The coverage resulting from this analysis pipeline is plotted in the left of figure \ref{CoverCalKiDS}. The red diagonal line indicates the perfect coverage for an unbiased analysis. Measured coverage probabilities above the red line indicate conservative credibility contours, which are strictly speaking too wide. Measured coverage probabilities below the red line indicate credibility contours which are too narrow. The purple data points depict the measured coverage with error bars. As can be seen, the posterior with the correct covariance matrix undercovers slightly, meaning it is slightly too narrow.   
This reflects that the adopted priors are slightly informative, as is well known in weak lensing \cite{ProbMatchPriors,KiDS,Maccrann}.

Next, we analyse the 100.000 simulations purposefully with a biased covariance matrix. We left- and right-multiply the KiDS covariance with a diagonal biasing matrix $\sfB$, whose diagonal elements are drawn from a Gaussian distribution 
\begin{equation}
\sfB = \mathrm{diag}(b_{11}, b_{22},...,b_{nn}), \ \ \mathrm{with} \ \ b_{ii} \sim \calG(1,\sigma^2).
\label{noise}
\end{equation}
The larger the standard deviation $\sigma$, the larger will be the bias in $\cinv_\mathrm{B} = \sfB^\top \cinv \sfB$. The relative difference between original covariance matrix, and biased covariance matrix then follows from the mean and standard deviations of $\sfB^\top \cinv \sfB$. Per matrix element, we have on average
\begin{equation}
\begin{aligned}
    \big\langle (\cinv_\mathrm{B})_{ij}  \big\rangle & = (\cinv)_{ij} \big\langle b_{ii} b_{jj} \big\rangle\\
                                  & = (\cinv)_{ij}
    \end{aligned}
    \label{mean}
\end{equation}
and using $\langle b_{ii}^2\rangle = \sigma^2 +1 \ \forall i $, the variance follows to be
\begin{equation}
    \begin{aligned}
   \mathrm{Var}[(\cinv_\mathrm{B})_{ij}] & = \big\langle [ (\cinv)_{ij} ]^2 \big\rangle -(\cinv)^2_{ij} \\
   & =  (\cinv)^2_{ij} [\sigma^4 + 2\sigma^2].
    \end{aligned}
    \label{stdev}
\end{equation}
According to equation~\ref{mean} the bias vanishes on average, and has according to equation \ref{stdev} a standard deviation of $s = \cinv_{ij} \sqrt{\sigma^4 + 2\sigma^2}$.
The relative difference $r$ between biased and correct covariance matrix is then
\begin{equation}
    r = \sqrt{\sigma^4 + 2\sigma^2},
\end{equation}
which is independent of matrix indices $ij$. The relative differences $r$ can be compared to the literature: for example, the DES reanalysis of the KiDS-450 data \cite{DESdoesKiDS} implemented 40 percent changes in the covariance matrix elements (see Figure 1 of \cite{DESdoesKiDS}). Current approximate covariance matrices in weak lensing are therefore uncertain to approximately a degree of $r \approx 0.4$.

We study such example biases in figure \ref{CoverCalKiDS}, for $r = 0.07, r=0.14$ and $r = 0.36$. Analyzing the data with such biased covariance matrices causes the posterior to preferentially peak in the wrong region of parameter space, thereby excluding the true cosmology too often. The left panel of figure \ref{CoverCalKiDS} illustrates this effect by showing how the best-fitting cosmologies are shifted. To each of these new best-fitting cosmologies belongs a new posterior (not plotted) whose credibility contours are of approximately the same shape as those of the original posterior, only centered on the new best fits. The blue and pink coverage measurements in the right panel of figure \ref{CoverCalKiDS} indicate how quickly the posterior begins to undercover if the biases of such covariance matrices are not mitigated. 

Since the left of figure \ref{CoverCalKiDS} indicates that for current levels of covariance matrix uncertainty ($r = 0.36)$ the best fit scatters over nearly the entire undebiased posterior, we conclude that such uncertainties definitely need to be propagated. We illustrate such a propagation first for the traditional Bayesian marginalization, and then for coverage correction.

Given our bias model with $\sfB$, the posterior of cosmological parameters when marginalized over $\sfB$ is given by
\begin{equation}
\calP(\both|\fatx) = \int \calG\left(\bomu(\both),\fatx,\sfB^T\cinv\sfB\right)\pi(\sfB) \mathd\sfB,
\label{margin}
\end{equation}
where the uncertainty of $\sfB$ is
\begin{equation}
\pi(\sfB) = \prod_i \calG(1,\sigma^2).
\end{equation}
The matrix-variate integration $\mathd \sfB$ is element-wise which becomes quickly numerically prohibitive due to the curse of dimensionality. For the 130-dimensional diagonal $\sfB$ used in equation \ref{noise}, it is still feasible, and we implement it via a Monte-Carlo integration. The resulting posterior, $\calP(\both|\fatx)$, is depicted in figure \ref{Coverplot} in blue dotted contours, and is wider than the original KiDS posterior (solid grey contours) due to the marginalization.

The Bayesian marginalization was here only possible since we knew the model which caused the bias. In a realistic analysis, such a model is not known, and we need to propagate the bias blindly via coverage calibration.

We therefore compute the posterior
\begin{equation}
\calP(\both|\fatx,\sfB^T\cinv\sfB),
\end{equation}
of which we know that it must be biased to unknown degree, due to having used the covariance matrix of unknown bias. The measured coverages in figure \ref{CoverCalKiDS} reveal that for $\sigma=0.25 (r=0.36)$, the credibility contour which contains 68\% of the posterior volume, only covers the true cosmology 42\% of the time. In contrast, the contour which contains 92\% of the posterior volume, included the true cosmology 68\% of the time. The 92\% credibility contour of the biased posterior is hence only the 68\% credibility contour after debiasing. Figure \ref{Coverplot} shows that this coverage calibration complies with the Bayesian marginalization, with the advantage that it required no parametric model.

\subsection{Mitigating redshift uncertainties by coverage calibration}
\label{debred}
In this section, the aim is to propagate redshift uncertainties in a non-parametric manner, for the following reasons.

Estimating the redshift $z$ of a galaxy becomes difficult when only photometric flux measurements are available. Tomographic weak lensing analyses assign galaxies to distributions $n_i(z)$, where $i$ denotes the bin, $i \in [1,r]$. If the redshifts have to be determined photometrically, then the estimated galaxy distributions are uncertain and we write $\hat{n}_i(z)$. There will thus exist a probability distribution
\begin{equation}
    \calP( \{\hat{n}_i(z) \} |\{n_i(z) \}),
    \label{zdist}
\end{equation}
where the curly braces indicate the set of all tomographic bins.

\begin{figure*}
\includegraphics[width=0.99\textwidth]{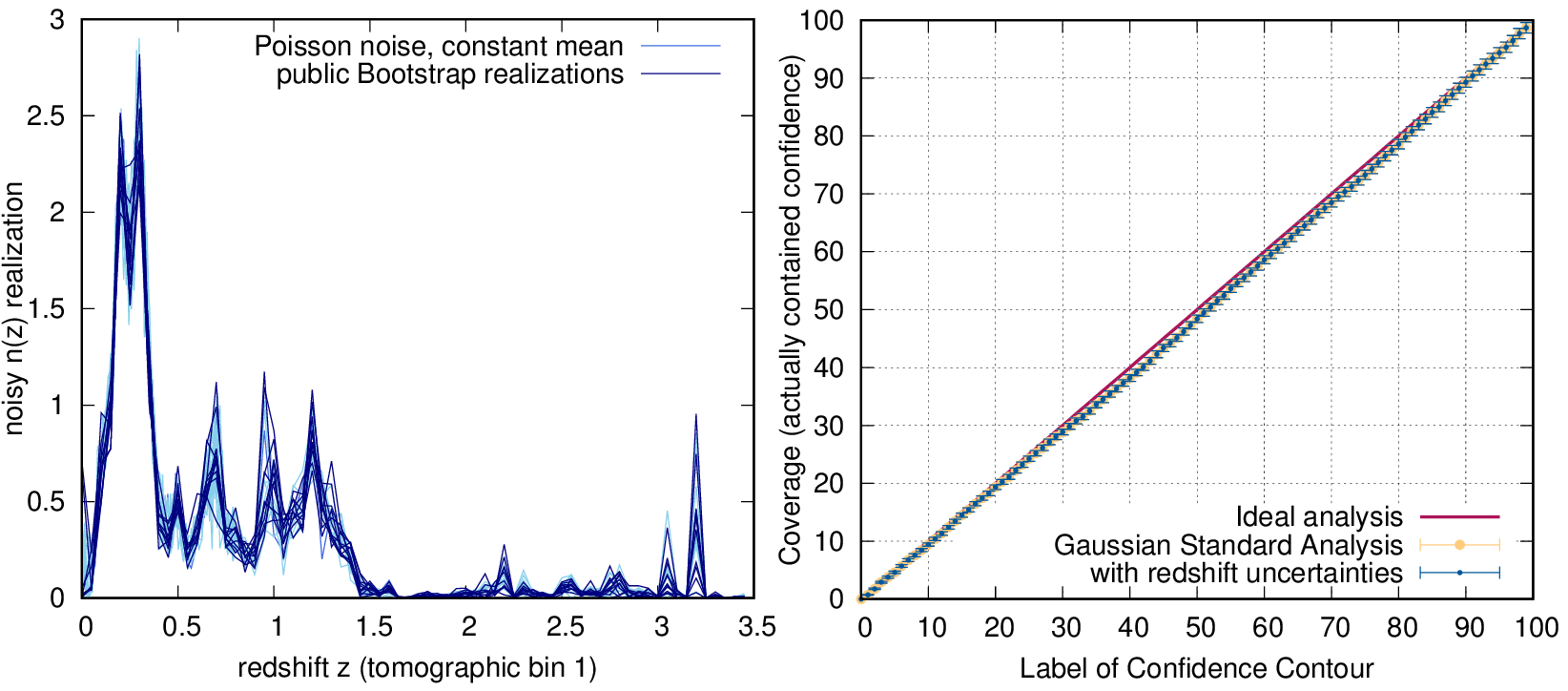} 
\caption{Left: Examples of random realizations of noisy redshift distributions, here for the first tomographic redshift bin from KiDS-450. Right: The coverage is essentially unaffected by redshift noise, illustrating that redshift noise is in current analyses sub-dominant to cosmic variance and shape noise.}
\label{Redshifts}
\end{figure*}

Propagating uncertainty from the $\hat{n}_i(z)$ through weak lensing analyses is difficult. Bayesian analyses would try to establish the precise functional form of $\calP$ in equation \ref{zdist}, and then marginalize over it, resulting in the marginal posterior of cosmological parameters $\both$
\begin{equation}
\calP(\both|\boldsymbol{\xi}_\pm) = \int \calP\left( \both,\boldsymbol{\xi}_\pm,\{\hat{n}_i(z)\},\boC \right) \calP( \{\hat{n}_i(z) \} |\{n_i(z) \}) \ \mathd n_1 ... \mathd n_r.
\label{redmarg}
\end{equation}
This integral is numerically extremely costly\footnote{Already figure \ref{Redshifts} required a CPU-time of 26 days in parallel on 10 modern Xenon CPUs (100 times as long as a computation for a single redshift realization).}, and has to date not yet been solved. Consequently, the current standard approach is to introduce nuisance parameters instead. 

Both KiDS and DES introduce nuisance parameters which shift the centers of redshift bins \citep{VanUitert,Troxel}. It has however been found \cite{VanUitert}, that the nuisance parameter originally introduced for the intrinsic alignment amplitude also fits to redshift uncertainties. This problem occurs because nuisance parameters \emph{fit}, i.e.~they are part of an inverse problem which enables them to compensate for unintended systematics. Note also, that the adopted parametric nuisance model is limited in the sense of not being able to create deformations of $\hat{n}_i(z)$ which leave the central redshifts invariant.

We therefore wish to study the impact of redshift uncertaintites in isolation. Consequently, we replace the nuisance parameters by a forward model of redshift noise. Since the forward model generates redshift noise only, a confusion with intrinsic alignments is excluded. We then use coverage calibration to propagate the redshift uncertainties into the cosmological parameters.

To implement $\calP( \{\hat{n}_i(z) \} |\{n_i(z) \})$, we use the published redshift uncertainties from KiDS-450. We use the weighted direct calibration `DIR' setup of KiDS, which matches spectroscopic galaxy observations and galaxies seen in KiDS. On average, DIR causes approximately $20\%$ uncertainties in each point, but we use the exact errors per point.

We implement four different forward models for redshift uncertainties. The first model generates  functions $\hat{n}_i(z)$ whose shape and mean vary. The second model varies the shape only, but keeps the central redshift fixed. This generates uncertainties which cannot be modelled by marginalizing over the mean redshift. For both cases, we use two noise processes: Poisson realizations and the public Bootstrap realizations from KiDS \citep{KiDS}.

Examples of the resulting noisy redshift distributions are shown in the left panel of figure \ref{Redshifts}. For each of these, we compute the theoretical prediction for the KiDS-450 data vector, and analyze it with a Gaussian likelihood, using the public KiDS-covariance matrix. The right panel of Figure \ref{Redshifts} reveals that none of the four noise models caused the posteriors to undercover -- this means that reported problems with redshifts in KiDS-450 must arise from a bias, or confusion with another systematic effect. Redshift noise in isolation, as here studied, seems to be subdominant to shape noise and cosmic variance, as included in the covariance matrix.

\section{Forecasts for dark energy constraints with a Euclid-like survey}
\label{Euclid}

As the precision of cosmic surveys improves, the relative impact of formerly negligible biases increases.
The upcoming Euclid survey \cite{DefStudyRep}, but also its sibling surveys LSST and WFIRST \cite{Synergies}, will study the cosmological standard model, and its constituents.
The cosmological standard model $\Lambda$CDM is based on a cosmological constant $\Lambda$ and cold dark matter (CDM). In $\Lambda$CDM, $\Lambda$ takes the role of dark energy, and physics beyond the standard model accordingly often introduces additional parameters, $w_0$ and $w_a$, for extended dark energy phenomenology \cite{Class1,Class2}. In $\Lambda$CDM, these parameters take values $w_0 = -1$, and $w_a = 0$. If the upcoming Euclid analyses exclude this point with high significance, then $\Lambda$CDM is ruled out and a new standard model is needed -- or a bias occurred.

Due to the complexity of the data analysis, the occurrence of an unrecognized bias is of course possible, but our method is able to tell these biases and new physics apart.

We imagine a Euclid-like survey develops a likelihood, which is as accurate as possible, and which does not rely on simulations. If the likelihood is very accurate, then our method will need to correct only minor outstanding biases, resulting in a minor increase of credibility contours. This likelihood is then to be augmented by few, but very accurate, end-to-end simulations for $\Lambda$CDM. We here forecast the number of simulations needed to guarantee that $\Lambda$CDM is not discarded due to unrecognized biases.

\begin{table}
\centering
\caption{We list which posterior volume contours can be guaranteed to contain at least the probability stated, thereby guaranteeing the absence of biases in constraints on the dark energy parameters $w_0$ and $w_a$.
The number of simulations is $N$.  The percent probability that shall be guaranteed to be contained in a certain contour is listed in the table header. The body list which posterior volume contour is guaranteed to contain at least this probability, or more. If infinitely many accurate simulations are run, then posterior volume can be guaranteed to directly measure probability (first line). For fewer simulations (remaining lines), biases are possible, such that contours can only be guaranteed to contain a somewhat lower probability than their encased volume would suggest. As an example, for 225 simulations, a bias is either found and mitigated, or if no bias is found, then the 92.5\% posterior volume contour can be guaranteed to contain the true parameters at least 90\% of the time. Accordingly, if $\Lambda$CDM were excluded by the 92.5\% contour, then a new standard model might be considered. The last column indicates the percentual decrease of the figure of merit, as a function of number of simulations.}
\label{Tab1}
\begin{tabular}{|ll||l|l|l|l|} 
\hline
N & $\sqrt{N}$ &   $\geq 68\%$ probability &   $\geq 90\%$ probability &   $\geq 95\%$ probability & $\Delta$FoM\\
\hline 
$\infty$ & $\infty$ & 68\% volume cont. & 90.0\% volume cont. & 95.0\% volume cont. & 0\%\\
625 & 25       &    70\% volume cont.  &  91.5\% volume cont. &     96\% volume cont. & 6\%\\
400 & 20       &    71\% volume cont. &   92.0\% volume cont. &     96\% volume cont. & 8\%\\
225 & 15       &    72\% volume cont. &   92.5\% volume cont. &    96.5\% volume cont. & 10\%\\
100 & 10       &    73\% volume cont. &  93\% volume cont.&       97\% volume cont. & 12\%\\
49 &  7        &    75\% volume cont. &  94\% volume cont.&       97.5\% volume cont. & 17\%\\
25 & 5         &    77\% volume cont. &  95\% volume cont. &      98\%  volume cont. & 22\%\\
		\hline
	\end{tabular}
\end{table}

According to \cite{DefStudyRep}, Euclid's prime scientific target is the determination of the dark energy equation of state parameters $w_0$ and $w_a$ to a precision of 
\begin{equation}
\begin{aligned}
\sigma(w_0)  = 0.02, \ \ \sigma(w_a) = 0.1,\\
\end{aligned}
\end{equation}
where $\sigma$ is the 1-sigma standard deviation. In a Gaussian approximation, the joint confidence contours of $w_0$ and $w_a$ are elliptical, and the figure of merit (FoM) measures this ellipses area.

For $N$ simulations, our coverage estimator $\hat{C}_\alpha$ has a standard deviation of $\sigma = \sqrt{\alpha(1-\alpha)/N}$. For $N$ simulations, it will thus detect biases which change confidence contours by more than $\sigma$. It cannot detect biases which change the coverage by less than $\sigma$, and accordingly 
\begin{equation}
 \hat{C}_\alpha^{\rm low} =   \hat{C}_\alpha - \sigma,
\end{equation}
is a conservative lower estimate of the coverage probability, to be interpreted as the `most conservative scenario' of mitigating all possible biases which could not yet be ruled out. A credibility contour which reaches coverage $\hat{C}_\alpha^{\rm low}$ contains the true parameters \emph{at least} with probability $\alpha$, and likely more.

\begin{figure}
\centering
\includegraphics[width=0.99\textwidth]{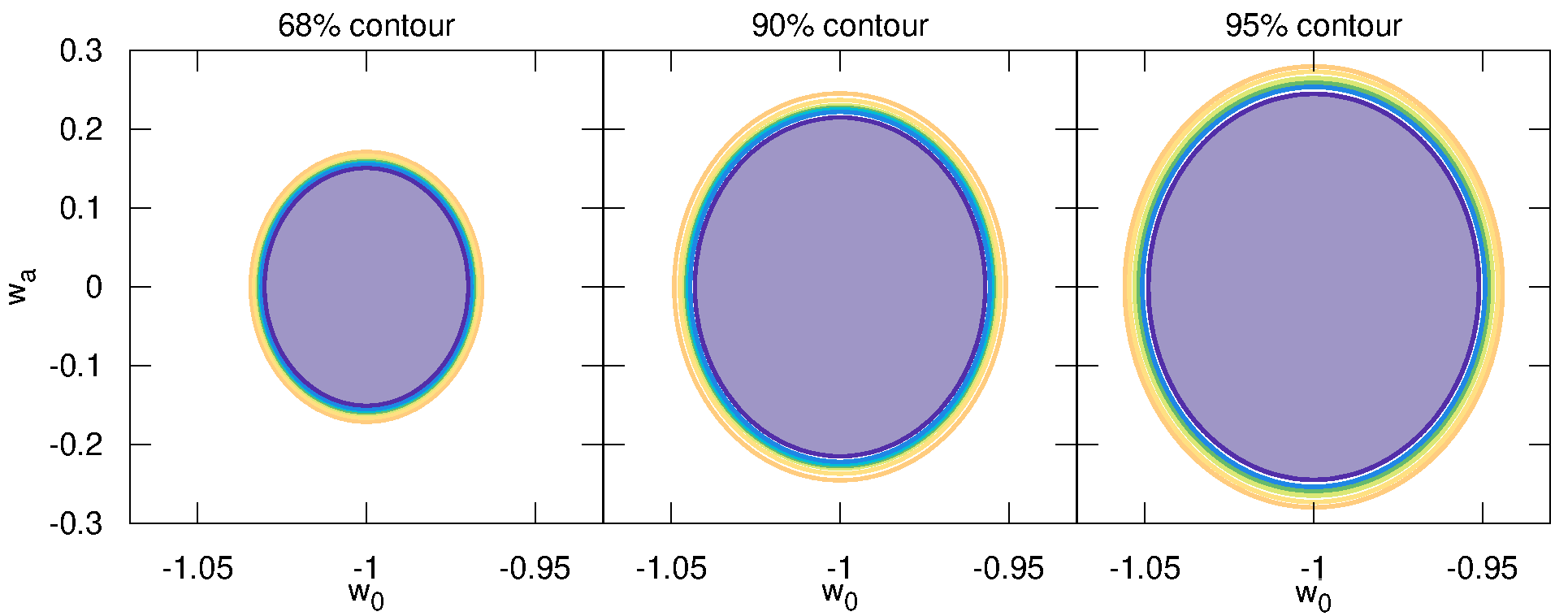} 
\caption{Forecast of conservative credibility contours for the dark energy parameters $w_0,w_a$. Filled inner contour: Euclid's nominal precision, corresponding to infinitely many simulations and a bias-free inference. Open contours: conservative contours, ranging from 625 (blue) to 25 simulations (orange), spaced as in table \ref{Tab1}. Right panel: If the Euclid analysis is augmented by 25 accurate simulations (orange), and no bias needs to be corrected, then its here shown conservative $\geq$95\% credibility constraints on $w_0$ are still approximately 25 times more constraining than current KiDS or DES analyses.}
\label{fig:Euclid}
\end{figure}

 In figure~\ref{fig:Euclid} we forecast such conservative credibility contours for Euclid. The inner filled contour depicts the 68, 90, and 95 percent credibility contours for Euclid at its nominal precision. These correspond to a bias-free inference with infinitely many simulations. If fewer simulations are available, then the conservative contours increase in size, which is depicted in open contours as a function of $N$, as given in table \ref{Tab1}.

 The right panel of figure \ref{fig:Euclid} can be compared to current constraints from DES and KiDS: both of these surveys measure an equation of state $w_0$, but keeping $w_a$ fixed to its fiducial value. Both surveys currently achieve approximately $-2.0 < w_0 < -0.4$ \cite{KiDS, Troxel}. Figure \ref{fig:Euclid} therefore indicates that if Euclid's data can be augmented by 25 simulations, then either biases can be detected and mitigated, or if no biases are detected but the shown conservative contours are chosen, then Euclid will still achieve approximately 25 times the precision of KiDS and DES for $w_0$.

\section{Discussion}
This paper presented a method to mitigate biases, recognized or unrecognized, even when a Bayesian solution cannot be conducted. Our method takes any existing data analysis pipeline as input, and runs it on simulations and the real data alike. It then measures the \emph{coverage probability} of credibility contours and corrects for it, if found to be off. This produces debiased contours as particle physicists (and many cosmologists) expect them to be: under a repetition of the experiment, the 68 percent confidence contour will contain the true parameters 68 percent of the time, despite the data being analyzed with an imperfect pipeline. Our method can also be understood as a sanity check for any cosmological analysis.

We showed how approximate covariance matrices determine where a likelihood peaks, and that a reduced-$\chi^2$ of order unity does not indicate an unbiased best-fit. To illustrate both points, we forced the original KiDS-450 data set to peak at the Planck best-fitting cosmology, with the exact same $\chi^2$. We then used our method to show how inferences with approximate covariance matrices can be debiased.

We also isolated the impact of uncertain redshifts by using a forward model, since an inverse treatment was found to confuse redshift uncertainties and intrinsic alignments \citep{VanUitert}. We found that in isolation, current redshift uncertainties are fully subdominant to shape noise and cosmic variance in current weak lensing analyses. Our study focuses on \emph{uncertaintites} not \emph{biases} in redshifts.

Finally, we illustrated that a pessimistic analysis of Euclid-like data, will very likely constrain the dark energy equation of state parameters by a factor of at least 25 better than current KiDS and DES analyses. This statement assumes that 25 end-to-end simulations of Euclid-like data can be provided alongside an independent likelihood.

This paper, in conjunction with \cite{SHInsuff}, now found repeatedly that \emph{data vector truncation} influences cosmological parameter constraints: the problematic data points always occured at the extreme of  angular ranges. This motivates that blinding strategies should be kept for all upcoming surveys.

Our method is applicable to many more examples, and the code is hence public at github.com/elenasellentin/Mitigate\_Unrecognized\_Biases.

\section*{Acknowledgements}
We appreciate the public data products of the KiDS consortium, without which this research would not have been possible. It is a pleasure to thank Ruth Durrer and Catherine Heymans for scientific discussions and long-term support.

\bibliography{TDist}
\bibliographystyle{JHEP}
\end{document}